\documentstyle[prl,aps,preprint]{revtex}
\begin{document}
\renewcommand{\theequation}{\arabic{section}.\arabic{equation}}

%\draft
%\twocolumn[\hsize\textwidth\columnwidth\hsize\csname @twocolumnfalse\endcsname
\title{
Interactions between Membrane Inclusions on Fluctuating Membranes
}
\author{Jeong-Man Park and T. C. Lubensky}
\address{
Department of Physics, University of Pennsylvania, \\
Philadelphia, PA 19104
}
\maketitle
 
\begin{abstract}
We model membrane proteins as anisotropic objects
characterized by  symmetric-traceless tensors and
determine the coupling between these order-parameters and membrane curvature.
We consider the interactions 1) between transmembrane proteins 
that respect up-down (reflection) symmetry of
bilayer membranes and that have circular or non-circular cross-sectional areas
in the tangent-plane of membranes, 2) between 
transmembrane proteins that break reflection symmetry and 
have circular or non-circular cross-sectional areas, and 
3) between non-transmembrane proteins.
Using a field theoretic approach, we find non-entropic $1/R^{4}$ interactions
between reflection-symmetry-breaking transmembrane proteins with circular 
cross-sectional area and entropic $1/R^{4}$ interactions between 
transmembrane proteins with circular cross-section that do not
break up-down symmetry in agreement with previous calculations.
We also find anisotropic $1/R^{4}$ interactions between 
reflection-symmetry-conserving transmembrane proteins with 
non-circular cross-section,
anisotropic $1/R^{2}$ interactions between reflection-symmetry-breaking
transmembrane proteins with
non-circular cross-section, and non-entropic $1/R^{4}$ many-particle
interactions among non-transmembrane proteins.
For large $R$, these interactions are considerably larger than Van der Waals
interactions or screened electrostatic interactions and might provide the 
dominant force inducing aggregation of the membrane proteins.
\end{abstract}
%]
%\clearpage
%\narrowtext
\par
\section{Introduction}
Recently, the structure and properties of model and biological membranes
have been studied extensively.
Biological membranes play a central role in both the structure and function
of cells.
%Biomembranes basically define compartments, each membrane associated with
%an inside and an outside.
%Also, they determine the nature of all communication between the inside and 
%outside.
%This may take the form of actual passage of ions or molecules betwen the two
%compartments or may be in the form of information, transmitted through
%conformational changes induced in membrane components.
%Model membranes, bilayers made from lipids and water, exhibit many of
%the attributes of biological membranes.
%For instance, bilayers can create separate compartments as in the case of
%cells, and can fuse as can biological membranes.
%Biomembranes provide boundaries that separate and determine all
%communication between the inside from the outside of cells.
Biomembranes divide living tissue into different compartments or cells
and act as cell boundaries.
They determine the nature of all communication between the inside and
the outside of cells.
This communication can take place via the actual passage of ions or molecules
between two compartments or via conformational changes induced in membrane
components.
Model bilayer lipid membranes in aqueous environments exhibit many of
the attributes of the biological membranes.
For example, these membranes can form vesicles or more complex structures
that divide space into separate compartments, which like cells can fuse
or divide.
However, there are many properties of biomembranes that cannot be mimicked by
lipid bilayers.
Energy-driven transport of ions across membranes and receptor-mediated events 
are only a few of the myriad of membrane-associated functions that 
lipid bilayers
are incapable of performing on their own.
%What is required is are membrane proteins.
%Thus, it is important to incorporate membrane inclusions, such as membrane 
%proteins, in model membranes to study the properties of biological membrane
%properly \cite{Warren,Voet,Alberts,Gennis,Tanford}.
Such processes are mediated by proteins that are attached to or dissolved
in biological membranes \cite{Warren,Voet,Alberts,Gennis,Tanford}.
Thus in order to make lipid bilayers more realistic models of biomembranes,
it is necessary to introduce into them membrane inclusions such as proteins.
\par
Membrane proteins are classified as integral proteins or peripheral proteins
according to how tightly they are associated with membranes. Integral
proteins are so tightly bound to membrane lipids by hydrophobic forces
that they can be freed only under denaturing conditions.
Peripheral proteins associate with a membrane by binding at 
the membrane  surface; they can be non-destructively dissociated from 
the membrane by relatively mild procedures. 
Some integral proteins, known as transmembrane proteins, span the membrane,
whereas others are attached to a specific surface of a membrane.
For brevity, we refer the latter proteins as non-transmembrane proteins 
to distinguish
these from transmembrane proteins.
However, no proteins are known to be completely buried in a 
membrane \cite{Voet}.
\par
%It has been proven very difficult to obtain detailed structural information
%about membrane proteins because membrane proteins are not amenable to
%normal sequencing techniques.
%The insoluble characteristics of most membrane proteins made primary structure
%determination difficult.
%Thus, progress on two membrane proteins, glycophorin and cytochrome $b_{5}$,
%provide the only models for most other membrane proteins 
%\cite{Voet,Alberts,Gennis}.
%Membrane proteins are generally viewed as being folded so as to present
%a nonpolar hydrophobic surface which can interact with the nonpolar portions
%of the lipid bilayer \cite{Gennis,Tanford}.
%Polar or charged regions of the protein can interact with the lipid 
%headgroups at the surface of the bilayer (Fig.~\ref{model-pr}a).
%\par
%Recent discoveries indicate how protein molecules can facilitate 
%the passage of hydrophilic substances into and out of the cell. One possible
%mechanism is based on a protein with a hydrophobic region on part 
%of its surface, but with hydrophilic amino acids covering the majority of  
%its surface.
%If several such proteins are present in the same membrane, stability is 
%increased if they aggregate so as to bring their hydrophilic faces together.
%A ring of three or more such proteins can form a hydrophilic channel through 
%the bilayer (Fig.~\ref{model-pr}b).
%\par
Interactions between membrane proteins is expected to be controlled by lipid
affinity, direct interactions such as electrostatic and Van der Waals 
interactions, and indirect interactions mediated by the 
membrane \cite{Israela}.
The latter interaction arises from thermally-driven undulations of the
membrane and is analogous to the Casimir force between conducting plates.
Since the degree to which fluctuations of a membrane are restricted
depends on distance between membrane proteins, its free energy also
depends on this distance, decreasing with decreasing separation. 
This implies an attractive force, which leads to a tendency for 
membrane proteins to aggregate.
This indirect interaction between membrane inclusions was
first calculated by Goulian, Bruinsma, and Pincus \cite{Goul}.
Before presenting the interaction models introduced in this paper in detail,
in Sec.II we will review its results, indicating where
they differ from and extend those of Goulian {\it et al.} 
In Sec.III, we introduce a phenomenological model 
(Model I) for protein interactions. 
In this model, proteins
are characterized by symmetric-traceless tensors depending on the cross-section
shapes of proteins on the membrane, and interactions are described by  
symmetry-allowed couplings between these tensor order-parameters and 
the curvature tensor of the fluctuating membrane.
In Sec.IV, we introduce another phenomenological model
(Model II), in which there is an
interaction between a membrane protein and membrane lipids at its perimeter. 
At the perimeter, lipids tend to align
with the direction of the protein at a certain angle depending on whether
or not proteins break the up-down symmetry of the bilayer membrane.
The interaction is described by the fluctuation of the normal vector of the
membrane around this preferred direction.
The interaction between non-transmembrane proteins is described in Sec.V.
Non-transmembrane proteins are proteins with preferred center-of-mass positions
not at the center of the bilayer membrane. We expand the potential energy
in terms of the deviation from this preferred position and include other
couplings considering the symmetry to calculate the interaction between 
non-transmembrane proteins.
Finally, a discussion is given in Sec.VI.
\setcounter{equation}{0}
\section{Review and Summary}
\subsection{Review of the previous work}
The interaction between membrane inclusions with circular
cross-section on the membrane has been calculated by
Goulian {\it et al.} \cite{Goul}.
They use the Helfrich-Canham Hamiltonian \cite{Helf,Canham},
\begin{equation}
{\cal H}_{0} = \frac{1}{2}\int d^{2}u\sqrt{g} (\kappa H^{2}
                                          + \bar{\kappa} K),
\end{equation}
to describe fluctuations of the membrane, 
where $\kappa$ and $\bar{\kappa}$ are the bending
and the Gaussian rigidities and $H$ and $K$ are the mean and the Gaussian 
curvatures, respectively.
The surface tension is not taken into account since it effectively
vanishes \cite{Brochard,David}.
Within inclusions with circular cross-section, the constants $\kappa$ and 
$\bar{\kappa}$ are assumed
to differ from those of the surrounding membranes.
For some inclusions, such as proteins, the circular regions are assumed to 
be rigid with $\kappa=-\bar{\kappa}=\infty$. 
This case is the strong-coupling regime. 
On the other hand, for regions with excess concentrations of lipids, 
$\kappa$ and $\bar{\kappa}$ within circular regions are assumed to have
values close to those of the surrounding membrane. 
This case is the perturbative regime.
In both the strong-coupling and perturbative regimes, Ref. \cite{Goul}
finds that there is an entropic 
$1/R^{4}$ interaction, which is proportional to the temperature and 
to the square
of the area of the circular region.
Also, at low temperatures, there is a non-entropic $1/R^{4}$
interaction between proteins which varies with the square of 
angle of contact between membrane and proteins.
Recently, Golestanian, Goulian, and Kardar \cite{Goul2}
extended the calculation in
Ref. \cite{Goul} to the interaction between two rods embedded in a fluctuating
membrane. They find an anisotropic $1/R^{4}$ interaction between
two rods.
\subsection{Summary of the present work}
\begin{figure}
\centerline{\psfig{figure=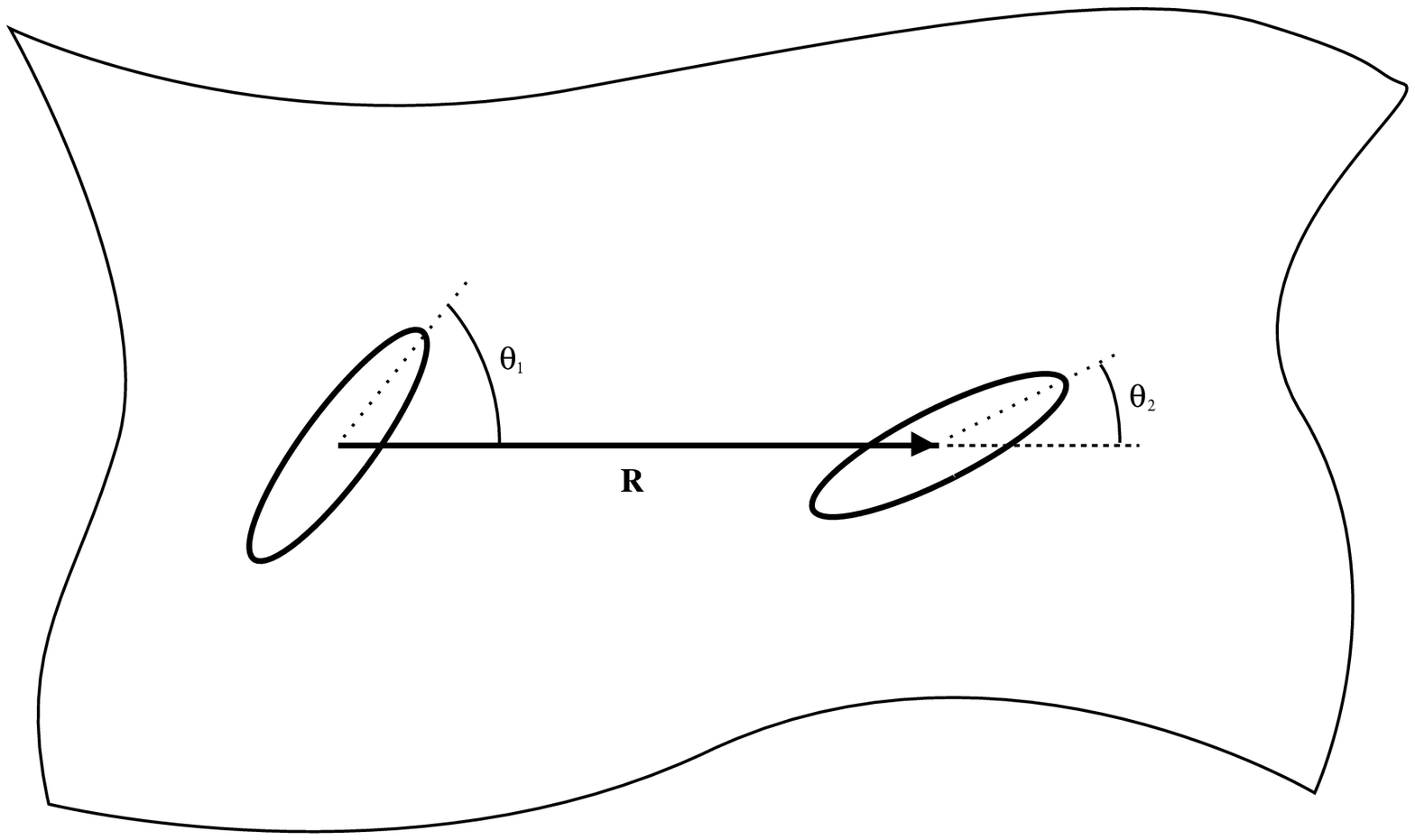}}
\caption{Proteins make angles $\theta_{i}$ measured with respect to
the separation vector ${\bf R}$. The distance between proteins $R$ is
taken to be much larger than protein size.
}
\label{fig1}
\end{figure}

In this paper, we introduce three models for the interaction between
membrane inclusions such as transmembrane proteins and non-transmembrane
proteins.
First, we present Model I. Here, proteins are characterized
by symmetric-traceless tensor order-parameters, and the coupling between
these order-parameters and membrane curvature is determined by symmetry 
and power counting.
If proteins don't respect up-down symmetry, we allow for couplings that
break up-down symmetry. Otherwise, we require up-down symmetry.
For proteins with circular cross-sectional area on the membrane and
preserving up-down symmetry, we find the leading distance-dependent
free energy,
\begin{equation}
{\cal F} = \frac{k_{B}T}{4\pi^{2}\kappa^{2}R^{4}}
\int_{D_{1}}d^{2}x\int_{D_{2}}d^{2}y
\left[ \delta\kappa(x)\delta\bar{\kappa}(y) + 
\delta\kappa(y)\delta\bar{\kappa}(x) \right],
\label{perturb}
\end{equation}
where the coupling constants are denoted in terms of the bending and the
Gaussian rigidity differences, $\delta\kappa(x)$ and $\delta\bar{\kappa}(x)$,
between proteins and surrounding membrane.
This corresponds to the perturbative regime of Ref. \cite{Goul},
whose results we reproduce.
For proteins with circular cross-sectional 
area, up-down-symmetry-breaking couplings do not affect the leading
contribution to the free energy, and the leading distance-dependent free
energy remains identical to Eq. (\ref{perturb}).
We also calculate interactions between proteins
with non-circular cross-sectional area.
When up-down symmetry is conserved, 
we find an interaction energy
\begin{eqnarray}
{\cal F} & = & -k_{B}T \frac{{\cal A}^{2}}{64\pi^{2}\kappa^{2}R^{4}} 
 \left( (q_{4}Q_{4}+q_{2}Q_{2}^{2})(q_{4}Q_{4}+q_{2}Q_{2}^{2}+d_{2}Q_{2}^{2})
 \cos 4(\theta_{1}+\theta_{2}) \right.  \nonumber  \\
 &  &  \left. + 2(q_{4}Q_{4}+q_{2}Q_{2}^{2}+d_{2}Q_{2}^{2})d_{2}Q_{2}^{2}
       \cos^{2}2\theta_{1}\cos^{2}2\theta_{2}
            + (q_{2}Q_{2}^{2}+d_{2}Q_{2}^{2})q_{2}Q_{2}^{2} \right),
\end{eqnarray}
where ${\cal A}$ is the cross-sectional area of proteins, 
$Q_{2}$ and $Q_{4}$ are
magnitudes of 2nd-rank and 4th-rank tensor order parameters measuring
orientational anisotropy, respectively, and
$\theta_{i}$ are the angles of the directions of proteins measured with respect
to the separation vector between proteins (See fig. \ref{fig1}).
This anisotropic $1/R^{4}$ interaction contains anisotropic
$\cos^{2}2\theta_{1}\cos^{2}2\theta_{2}$ interaction in addition to
anisotropic $\cos 4(\theta_{1}+\theta_{2})$ interaction
also found in the recent independent work 
by Golestanian {\it et al.} \cite{Goul2}.
However, when up-down symmetry is broken, there is an 
anisotropic $1/R^{2}$ interaction:
\begin{equation}
{\cal F} = -\frac{{\cal A}^{2}d_{1}^{2}Q_{2}^{2}}{16\pi \kappa R^{2}}
\cos 2(\theta_{1}+\theta_{2}).
\end{equation}
Thus the leading term in the free energy falls off with 
separation as $1/R^{2}$ rather than $1/R^{4}$.
\par
\begin{figure}
\centerline{\psfig{figure=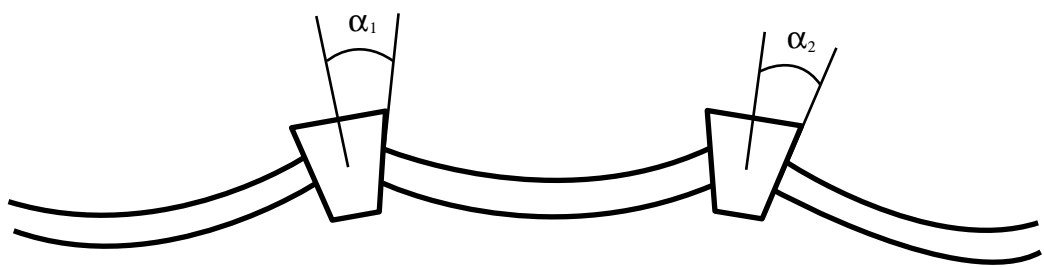}}
\caption{Contact angle $\alpha_{i}$ is measured between the direction of
$i$-th protein and the unit normal of the membrane at protein's perimeter.
}
\label{fig2}
\end{figure}

Next, we introduce Model II in which we impose a certain
boundary condition at the perimeter of proteins with the circular 
cross-sectional area. For proteins with up-down symmetry, we find
\begin{equation}
{\cal F} = -k_{B}T\frac{6{\cal A}^{2}}{\pi^{2}R^{4}}.
\label{micro-ud}
\end{equation}
This free energy looks similar to that of Ref. \cite{Goul} 
in the strong-coupling regime. 
When proteins break up-down symmetry, we find 
\begin{equation}
{\cal F} = -k_{B}T\frac{6{\cal A}^{2}}{\pi^{2}R^{4}} +
\frac{4\kappa {\cal A}^{2}}{\pi}
\frac{\alpha_{1}^{2}+\alpha_{2}^{2}}{R^{4}},
\end{equation}
where $\alpha_{i}$ is the contact angle between the direction of $i$-th protein
and the unit normal of the membrane (See fig. \ref{fig2}).
In the limit $T \rightarrow 0$, this free energy becomes the result in 
Ref. \cite{Goul} for the low-temperature regime. 
\par
Finally, we introduce a height-displacement model in which protein positions
normal to the membrane can vary. In this model, we find three- and four-body
interactions in addition to a two-body interaction.
These three- and four-body interactions also fall off as $1/R^{4}$ 
and are the same order of magnitude as two-body interaction.
\par
Consequently, by introducing three models to describe the interaction 
between membrane inclusions such as proteins, we recover all the 
results in Ref. \cite{Goul}. Furthermore, we obtain anisotropic 
interactions between proteins with non-circular cross-sectional area.
Also, we extend the calculation to the up-down symmetry breaking proteins
with non-circular cross-section and find anisotropic $1/R^{2}$ interaction
between them.
Moreover, using a height-displacement model, we find three- and four-body 
$1/R^{4}$ interaction in addition to two-body $1/R^{4}$ interaction.
\setcounter{equation}{0}
\section{Model I}
For a fluid membrane free of membrane proteins, the energy of membrane
conformations can be described by the Helfrich-Canham 
Hamiltonian \cite{Helf,Canham},
\begin{equation}
{\cal H}_{0} = \frac{1}{2}\int d^{2}u\sqrt{g} (\kappa H^{2}
                                          + \bar{\kappa} K),
\end{equation}
expressed in terms of the local mean and Gaussian curvatures.
We will work at length scales large compared with the membrane thickness but
small compared with the membrane's persistence length. 
Thus, we can parameterize the membrane in the 
Monge gauge ${\bf R} = ({\bf x},h({\bf x}))$ where ${\bf x} = (u_{1},u_{2})$.
In terms of ${\bf R}$ and the unit normal vector of the membrane ${\bf N}$,
the metric tensor $g_{\alpha\beta}$ is given by $\partial_{\alpha}{\bf R}\cdot
\partial_{\beta}{\bf R}$ and the curvature tensor $K_{\alpha\beta}$ is given
by ${\bf N}\cdot D_{\alpha}D_{\beta}{\bf R}$, where $D_{\alpha}$ is the 
covariant derivative along $u_{\alpha}$ direction on the membrane.
In the Monge gauge, 
\begin{equation}
H = \frac{1}{2}g^{\alpha\beta}K_{\alpha\beta} = \nabla^{2}h + O(h^{2}),
\end{equation}
\begin{equation}
K = \det g^{\alpha\beta}K_{\beta\gamma} = \nabla^{2}h\nabla^{2}h-
\partial_{\alpha}\partial_{\beta}h\partial_{\alpha}\partial_{\beta}h 
+ O(h^{4}).
\end{equation}
When the topology of the membrane is fixed, the Gaussian curvature term can
be dropped, and the leading term in 
${\cal H}_{0}$ in an expansion in derivatives of $h$ is
\begin{equation}
{\cal H}_{0} = \frac{1}{2}\kappa\int d^{2}x \nabla^{2}h \nabla^{2}h.
\end{equation}
Now let us consider the coupling between membrane proteins
and membranes. 
\subsection{Proteins with circular cross-section}
Membrane proteins can have arbitrary shapes; as a result their tangent-plane
cross-sections can be any shape.
Now we will compute the undulation mediated force between proteins
separated by a distance larger than the size of proteins.
For simplicity, let us first consider membrane proteins that have 
a circular cross-sectional area on the membrane. 
These proteins may be described by a scalar
density, $\rho$, which may be interpreted as the distribution
function of proteins 
describing the positions of proteins and the configurations of
protein's amino acid sequence
\begin{equation}
\rho({\bf x}) = \sum_{i} \frac{1}{\sqrt{g}}f_{i}({\bf x}-{\bf x}_{i}), 
\end{equation}
where ${\bf x}=(u^{1},u^{2})$ is a point on the membrane and 
the sum is over all proteins. The specific form of
$f_{i}({\bf x}-{\bf x}_{i})$ depends on the specific conformation of $i$-th
protein at the position ${\bf x}_{i}$.
It vanishes outside the protein cross section:
\begin{equation}
f_{i}({\bf x}) = \left\{ \begin{array}{ll}
                         f(x), & |{\bf x}| < a_{p}, \\
                          0  , & |{\bf x}| > a_{p}. 
                         \end{array}
                 \right.
\end{equation}
where $a_{p}=\sqrt{{\cal A}/\pi}$ 
is the radius of the protein where ${\cal A}$ is its cross-sectional area.
We assume all proteins are identical so that they are all described by the
the same function $f(|{\bf x}-{\bf x}_{i}|)$.
For membrane proteins, $a_{p}$ is
of order $10^{2}\AA$. If we model the protein as a uniform
cylinder, the distribution function of protein will be $f(x)=1$ inside
the projected area and $f(x)=0$ outside. In general, proteins have
non-uniform folding of the amino acid chain, and $f(x)$ will have small 
deviations from unity inside the circular cross-sectional region $D$.
In this case, we use
\begin{equation}
\int_{D} d^{2}x f(x) = {\cal A}
\end{equation}
as the definition of ${\cal A}$.
\par
When proteins do not break up-down symmetry, the relevant coupling
between $\rho$ and the height fluctuation field of the membrane
is
\begin{eqnarray}
{\cal H}_{\rm int} & = & \frac{1}{2}\int d^{2}x 
      [\alpha\rho({\bf x})K_{a}^{a}K_{b}^{b} + 
       \gamma\rho({\bf x})K_{b}^{a}K_{a}^{b}] 
      \nonumber   \\
   & = & \frac{1}{2}\sum_{i}\int_{D_{i}} d^{2}x
      [\alpha f(|{\bf x}-{\bf x}_{i}|) K^{a}_{a}K^{b}_{b} 
      + \gamma f(|{\bf x}-{\bf x}_{i}|) K^{a}_{b}K^{b}_{a}],
   \label{pm-coup}
\end{eqnarray}
where $D_{i}, i=1,2,\cdots$ denote circular regions occupied by membrane 
proteins.
The coupling constants $\alpha$ and $\gamma$ describe couplings between the
density inside protein's cross section and the curvature of a membrane. 
Thus these can be related to the bending and Gaussian rigidities:
\begin{equation}
\alpha f(x) = \delta\kappa(x) + \delta\bar{\kappa}(x) \;\;
, \;\;\; \gamma f(x) = -\delta\bar{\kappa}(x) \; ,
\label{alpha-gamma}
\end{equation}
where $\delta\kappa$ and $\delta\bar{\kappa}$ can be 
interpreted as the changes 
in the bending and the Gaussian rigidities due to the existence of proteins
on the membrane.
In the Monge gauge, to lowest order in $h$
\begin{equation}
K^{a}_{a} = -\nabla^{2}h \;\; , \;\;\;
K^{a}_{b}K^{b}_{a} = \partial_{a}\partial_{b}h\partial_{a}\partial_{b}h,
\end{equation}
and the relevant coupling becomes
\begin{equation}
{\cal H}_{\rm int} = \frac{1}{2} \int d^{2}x
[\alpha\rho({\bf x})(\nabla^{2}h)^{2} + 
\gamma\rho({\bf x})\partial_{a}\partial_{b}h
\partial_{a}\partial_{b}h].
\end{equation}
The free energy is given by
\begin{eqnarray}
\exp[-\beta{\cal F}] & = & \int [{\cal D}h] \exp[-\frac{1}{2}
          \beta\kappa\int d^{2}x
          (\nabla^{2}h)^{2}   \nonumber   \\
    &   & - \frac{1}{2}\beta \int d^{2}x
    [\alpha\rho({\bf x})(\nabla^{2}h)^{2} + 
     \gamma\rho({\bf x})\partial_{a}\partial_{b}h
          \partial_{a}\partial_{b}h].
\end{eqnarray}
We can use the cumulant expansion to calculate this form of the free energy.
We write
\begin{equation}
e^{-\beta({\cal F}-{\cal F}_{0})} = \left\langle 
\exp[-\frac{1}{2}\beta \int d^{2}x
[\alpha\rho({\bf x})(\nabla^{2}h)^{2} + 
 \gamma\rho({\bf x})\partial_{a}\partial_{b}h
\partial_{a}\partial_{b}h]] \right\rangle_{0}
\label{eq:free-pro}
\end{equation}
where $\langle \rangle_{0}$ denotes the ensemble average over the fluid 
membrane Hamiltonian only and 
\begin{equation}
e^{-\beta{\cal F}_{0}} = \int [{\cal D}h] \exp[-\frac{1}{2}
\beta\kappa\int d^{2}x(\nabla^{2}h)^{2}].
\end{equation}
The cumulant expansion gives
\begin{eqnarray}
\langle e^{V} \rangle_{0} & = & 
             \langle 1+V+\frac{1}{2}V^{2}+\cdots \rangle_{0}  \nonumber  \\
  & = & \exp\left[\langle V \rangle_{0} + \frac{1}{2}[\langle V^{2} \rangle_{0}
          -\langle V \rangle_{0}^{2}] + {\cal O}(V^{3})\right].
\label{eq:cum-exp}
\end{eqnarray}
Plugging Eq. (\ref{eq:free-pro}) into the cumulant expansion 
Eq. (\ref{eq:cum-exp}) and keeping terms up to order $h^{4}$, we find the free 
energy
\begin{eqnarray}
-\beta({\cal F}-{\cal F}_{0}) & = & \left\langle 
-\frac{1}{2}\beta \int d^{2}x
[\alpha\rho({\bf x})(\nabla^{2}h)^{2} + 
 \gamma\rho({\bf x})\partial_{a}\partial_{b}h
\partial_{a}\partial_{b}h] \right\rangle_{0}   \nonumber   \\
     &   & + \frac{1}{2}\left\langle 
(-\frac{1}{2}\beta \int d^{2}x
[\alpha\rho({\bf x})(\nabla^{2}h)^{2} + 
 \gamma\rho({\bf x})\partial_{a}\partial_{b}h
\partial_{a}\partial_{b}h])^{2} \right\rangle_{0}   \nonumber   \\
     &   & - \frac{1}{2}\left\langle 
-\frac{1}{2}\beta \int d^{2}x
[\alpha\rho({\bf x})(\nabla^{2}h)^{2} + 
 \gamma\rho({\bf x})\partial_{a}\partial_{b}h
\partial_{a}\partial_{b}h] \right\rangle_{0}^{2}.
\label{eq:free-exp}
\end{eqnarray}
This can be expanded in terms of the height correlation function
$G_{hh}({\bf x}-{\bf y})$ and its derivatives.
The height correlation function in the real space is
\begin{eqnarray}
G_{hh}({\bf x}-{\bf y}) & = & \langle h({\bf x})h({\bf y}) \rangle_{0}
                               \nonumber \\
   & = & \int \frac{d^{2}p}{(2\pi)^{2}}\frac{d^{2}q}{(2\pi)^{2}}
         e^{i{\bf p}\cdot {\bf x} + i{\bf q}\cdot {\bf y}} \langle 
         h({\bf p})h({\bf q}) \rangle_{0} \nonumber \\
   & = & \int \frac{d^{2}p}{(2\pi)^{2}} 
         \frac{e^{i{\bf p}\cdot({\bf x}-{\bf y})}}
{\beta\kappa p^{4}} = \frac{1}{16\pi\beta\kappa}R^{2}\ln R^{2},
\end{eqnarray}
where we used $\langle h({\bf p})h({\bf q})\rangle_{0} = 
(2\pi)^{2}\delta({\bf p}+{\bf q})/\beta\kappa
p^{4}$ in momentum space and ${\bf x}-{\bf y}={\bf R}$.
Then, taking derivatives we find
\begin{eqnarray}
\langle \partial_{a}\partial_{b}h({\bf x})\partial_{i}\partial_{j}h({\bf y}) 
        \rangle_{0}
    & = & \partial_{a}\partial_{b}\partial_{i}\partial_{j}
        G_{hh}({\bf x}-{\bf y})          \nonumber  \\
   & = & \frac{1}{4\pi\beta\kappa R^{2}}\left[ (\delta_{ab}\delta_{ij} + 
   \delta_{ai}\delta_{bj} + \delta_{aj}\delta_{bi}) \right.  \nonumber  \\
 & & -2(\hat{R}_{a}\hat{R}_{b}\delta_{ij} + \hat{R}_{a}\hat{R}_{i}\delta_{bj} +
     \hat{R}_{a}\hat{R}_{j}\delta_{bi} + \hat{R}_{b}\hat{R}_{i}\delta_{aj} + 
     \hat{R}_{b}\hat{R}_{j}\delta_{ai} + \hat{R}_{i}\hat{R}_{j}\delta_{ab}) 
     \nonumber  \\
 & & \left. +8\hat{R}_{a}\hat{R}_{b}\hat{R}_{i}\hat{R}_{j} \right] \nonumber \\
   & \equiv & \frac{1}{4\pi\beta\kappa R^{2}} T_{abij}(\hat{{\bf R}}),
\end{eqnarray}
where $\hat{{\bf R}}={\bf R}/R$, $R=|{\bf R}|$.
We proceed to calculate the terms in Eq.~(\ref{eq:free-exp}).
We are only interested in terms that depend on the distance 
between membrane proteins. We can, therefore, drop the first and the last terms
in the {\small RHS} of Eq. (\ref{eq:free-exp}) since they do not depend
on distance:
\begin{eqnarray}
\left\langle \int d^{2}x \rho({\bf x}) \partial_{a}\partial_{b}h
       \partial_{a}\partial_{b}h \right\rangle_{0}   
  & = & \int d^{2}x \rho({\bf x}) 
       \nabla^{4}G_{hh}({\bf x}-{\bf y})|_{{\bf y}={\bf x}} \nonumber  \\
 & = & \int d^{2}x \rho(x) \int\frac{d^{2}p}{(2\pi)^{2}}
       \frac{1}{\beta\kappa}  
       = \sum_{i}\int_{D_{i}}d^{2}x f(x) \int\frac{d^{2}p}{(2\pi)^{2}}
       \frac{1}{\beta\kappa}  \nonumber  \\
 & = & \frac{N}{(2\pi)^{2}\beta\kappa}\pi a_{p}^{2} \pi \Lambda^{2} =
       \frac{N}{4\beta\kappa}(a_{p}\Lambda)^{2},  
\label{eq:drop}
\end{eqnarray}
and similarly
\begin{equation}
\left\langle \int d^{2}x \rho({\bf x}) (\nabla^{2}h)^{2} \right\rangle_{0} 
= \frac{N}{4\beta\kappa}(a_{p}\Lambda)^{2},  
\end{equation}
where $N$ is the number of proteins.
In Eq. (\ref{eq:drop}) we introduced the cut-off for the
height fluctuation field $\Lambda \sim 1/a_{p}$ where $a_{p}$ is the radius
of the protein. 
The second term gives contribution to the 
distance-dependent free energy:
\begin{eqnarray}
 &   & \frac{\beta^{2}}{8}\int d^{2}xd^{2}y\rho({\bf x})\rho({\bf y}) 
\left[ 4\alpha\gamma \partial_{a}\partial_{b}\nabla^{2}G_{hh}({\bf x}-{\bf y})
\partial_{a}\partial_{b}\nabla^{2}G_{hh}({\bf x}-{\bf y}) 
\right. \nonumber  \\
 &   & \left. + 2\gamma^{2}
\partial_{a}\partial_{b}\partial_{i}\partial_{j}G_{hh}({\bf x}-{\bf y})
\partial_{a}\partial_{b}\partial_{i}\partial_{j}G_{hh}({\bf x}-{\bf y})
\right]  \\
 & = & \frac{\beta^{2}}{8}\int d^{2}xd^{2}y\rho({\bf x})\rho({\bf y}) 
\frac{1}{(4\pi\beta\kappa R^{2})^{2}}
\left[ 4\alpha\gamma T_{abii}(\hat{\bf R})T_{abjj}(\hat{\bf R}) 
+ 2\gamma^{2} T_{abij}(\hat{\bf R})T_{abij}(\hat{\bf R})
\right], \nonumber
\end{eqnarray}
where we kept only the leading distance-dependent terms and
${\bf R}={\bf x}-{\bf y}$.
Thus, the leading distance-dependent free energy is given by
\begin{equation}
-\beta{\cal F} = \frac{1}{4} \int d^{2}x \int d^{2}y \rho({\bf x}) U(|{\bf x}-
{\bf y}|) \rho({\bf y}),
\end{equation}
where
\begin{equation}
U(|{\bf x}-{\bf y}|) = \frac{(\alpha\gamma+\gamma^{2})}
{(\pi\kappa)^{2}|{\bf x}-{\bf y}|^{4}}.
\end{equation}
For two proteins separated by a distance $R$, the leading $R$ dependence is
\begin{equation}
-\beta{\cal F} = \frac{1}{2\pi^{2}\kappa^{2}R^{4}}
\int_{D_{1}}d^{2}x\int_{D_{2}}d^{2}y
\left[ \alpha\gamma + \gamma^{2} \right] f(x)f(y).
\end{equation}
Relating the couplings $\alpha$ and $\gamma$ 
with the variations of the bending rigidity and the
Gaussian rigidity as in Eq. (\ref{alpha-gamma}),
we recover the result of Goulian {\it et al.} \cite{Goul}
\begin{equation}
{\cal F} = \frac{k_{B}T}{4\pi^{2}\kappa^{2}R^{4}}
\int_{D_{1}}d^{2}x\int_{D_{2}}d^{2}y
\left[ \delta\kappa(x)\delta\bar{\kappa}(y) + 
\delta\kappa(y)\delta\bar{\kappa}(x) \right].
\end{equation}
\par
If membrane proteins break up-down bilayer symmetry, there is
another possible relevant coupling,
\begin{equation}
{\cal H}_{\rm int} = \frac{1}{2} \lambda \int d^{2}x \rho({\bf x}) K^{a}_{a}
= \frac{1}{2}\lambda \sum_{i}\int_{D_{i}} d^{2}x 
f(|{\bf x}-{\bf x}_{i}|) K^{a}_{a}. 
\end{equation}
However, this term does not contribute to protein-protein interactions 
since the 
distance-dependent contribution vanishes as follows,
\begin{eqnarray}
\int d^{2}xd^{2}y \rho({\bf x})\rho({\bf y})\langle \nabla^{2}h({\bf x})
\nabla^{2}h({\bf y}) \rangle_{0}  
   & = & \int d^{2}xd^{2}y \rho({\bf x})\rho({\bf y})
         \nabla^{4}G_{hh}({\bf x}-{\bf y})   \\
   & = & \frac{2}{\beta\kappa}
\int_{D_{1}}d^{2}x\int_{D_{2}}d^{2}y f(x)f(y)\delta(x-y) = 0. \nonumber
\end{eqnarray}
\subsection{Proteins with non-circular cross-sections}
So far we have, for simplicity, considered  protein-protein 
interactions when proteins have circular cross section.
However, in general proteins have asymmetric conformations giving
rise to non-circular foot prints on the membrane surface.
They can then be characterized by symmetric-traceless tensor order-parameters
such as $\tilde{Q}^{ab}, \tilde{Q}^{abcd}$, and so on:
\begin{equation}
\tilde{Q}^{ab}({\bf x}) = \sum_{\mu} \frac{1}{\sqrt{g}}
Q^{ab}_{\mu} f({\bf x}-{\bf x}_{\mu}),
\;\;\; \tilde{Q}^{abcd}({\bf x}) = \sum_{\mu} \frac{1}{\sqrt{g}}Q^{abcd}_{\mu}
f({\bf x}-{\bf x}_{\mu}),
\end{equation}
where ${\bf x}_{\mu}$ denotes the position of $\mu$-th protein and
$Q^{ab}_{\mu}$ and $Q^{abcd}_{\mu}$ are the symmetric-traceless tensors
constructed from the characteristic direction vector of $\mu$-th protein
on the membrane.
\par
When up-down symmetry is not broken, the relevent
coupling between inclusions and curvature is
\begin{equation}
{\cal H}_{\rm int} = \frac{1}{2} \int d^{2}x
\tilde{S}^{abcd}({\bf x}) K_{ab}K_{cd},
\end{equation}
where
\begin{equation}
\tilde{S}^{abcd}({\bf x})=q_{4}\tilde{Q}^{abcd}({\bf x})+
q_{2}\tilde{Q}^{ab}({\bf x})\tilde{Q}^{cd}({\bf x})+
d_{2}\tilde{Q}^{ac}({\bf x})\tilde{Q}^{bd}({\bf x}).
\end{equation}
The coupling constants $q_{4}$, $q_{2}$ and $d_{2}$ describe 
couplings between the protein tensor order parameters and 
the curvature of a membrane.
Results for membranes with circular cross-sections can be obtained by choosing
\begin{equation}
S^{abcd}_{\mu} = \alpha\delta^{ab}\delta^{cd} 
+ \gamma\delta^{ac}\delta^{bd},
\end{equation}
rather than insisting the order parameters be symmetric and traceless.
With this coupling, we proceed as before using the cumulant expansion.
For two proteins separated by a distance vector ${\bf R}$
from one to the other, the free energy becomes
\begin{eqnarray}
-\beta{\cal F} & = & \frac{\beta^{2}}{8} \int d^{2}x
          \int d^{2}y \tilde{S}^{abcd}({\bf x})
                      \tilde{S}^{ijkl}({\bf y})   
        \langle \partial_{a}\partial_{b}h({\bf x})
        \partial_{c}\partial_{d}h({\bf x}) \partial_{i}\partial_{j}h({\bf y}) 
        \partial_{k}\partial_{l}h({\bf y}) \rangle_{0}    \\
   & = &  \frac{\beta^{2}}{8} \int d^{2}x
          \int d^{2}y \tilde{S}^{abcd}({\bf x})
                      \tilde{S}^{ijkl}({\bf y})   
   \partial_{a}\partial_{b}\partial_{i}\partial_{j}G_{hh}({\bf x}-{\bf y})
   \partial_{a}\partial_{b}\partial_{i}\partial_{j}G_{hh}({\bf x}-{\bf y}).
\nonumber
\end{eqnarray}
In terms of the tensor $T_{abij}$ introduced before,
the final form for the free energy writes as
\begin{equation}
-\beta{\cal F} = \frac{1}{4}\int d^{2}x \int d^{2}y \tilde{S}^{abcd}({\bf x})
U_{abcd,ijkl}(|{\bf x}-{\bf y}|) \tilde{S}^{ijkl}({\bf y}),
\end{equation}
where
\begin{equation}
U_{abcd,ijkl}(|{\bf x}-{\bf y}|) = 
\frac{(T_{abij}(\hat{\bf R})T_{cdkl}(\hat{\bf R})+
T_{abkl}(\hat{\bf R})T_{cdij}(\hat{\bf R}))}
{8\pi^{2}\kappa^{2} |{\bf x}-{\bf y}|^{4}}.
\end{equation}
For two identical proteins separated by ${\bf R}={\bf x}_{1}-{\bf x}_{2}$, 
the leading distance-dependent free energy is found to be
\begin{equation}
{\cal F} = -k_{B}T 
\frac{{\cal A}^{2}S^{abcd}({\bf x}_{1})S^{ijkl}({\bf x}_{2})}
{64\pi^{2}\kappa^{2}R^{4}}
[T_{abij}(\hat{\bf R})T_{cdkl}(\hat{\bf R}) + 
T_{abkl}(\hat{\bf R})T_{cdij}(\hat{\bf R})].
\end{equation}
Now the free energy is anisotropic, depending on the direction of the
separation vector ${\bf R}$
and the orientation of proteins described by $S^{abcd}({\bf x}_{i})$.
$Q^{abcd}$ is a 4th-rank symmetric-traceless tensor, which can be expressed
as
\begin{eqnarray}
Q^{abcd} & = & Q_{4} \left( e^{a}e^{b}e^{c}e^{d} \right.  \nonumber   \\
   &   & - \frac{1}{6}(e^{a}e^{b}\delta^{cd} + e^{a}e^{c}\delta^{bd} +
e^{a}e^{d}\delta^{bc} + e^{b}e^{c}\delta^{ad} + e^{b}e^{d}\delta^{ac} +
e^{c}e^{d}\delta^{ab})  \nonumber   \\
   &   & \left. + \frac{1}{24}(\delta^{ab}\delta^{cd} + 
\delta^{ac}\delta^{bd} + \delta^{ad}\delta^{bc} ) \right),
\end{eqnarray}
and $Q^{ab}$ is a 2nd-rank symmetric-traceless tensor;
\begin{equation}
Q^{ab} = Q_{2} \left( e^{a}e^{b} -\frac{1}{2}\delta^{ab} \right),
\end{equation}
where $e^{1} = \cos\theta$ and $e^{2}=\sin\theta$ characterize the direction
of protein with $\theta$ measured with respect to the separation vector
${\bf R}$ and $Q_{2}$ and $Q_{4}$ are magnitudes of 2-fold and 4-fold 
anisotropy, respectively. 
Then, the free energy becomes
\begin{eqnarray}
{\cal F} & = & -k_{B}T \frac{{\cal A}^{2}}{64\pi^{2}\kappa^{2}R^{4}} 
 \left( (q_{4}Q_{4}+q_{2}Q_{2}^{2})(q_{4}Q_{4}+q_{2}Q_{2}^{2}+d_{2}Q_{2}^{2})
 \cos 4(\theta_{1}+\theta_{2})\right.  \nonumber  \\
 &  &  \left. + 2(q_{4}Q_{4}+q_{2}Q_{2}^{2}+d_{2}Q_{2}^{2})d_{2}Q_{2}^{2}
       \cos^{2}2\theta_{1}\cos^{2}2\theta_{2}
            + (q_{2}Q_{2}^{2}+d_{2}Q_{2}^{2})q_{2}Q_{2}^{2} \right),
\label{quad-int}
\end{eqnarray}
\par
Again, for proteins breaking up-down symmetry, we have the additional 
relevant coupling
\begin{equation}
{\cal H}_{\rm int} = \frac{1}{2} \int d^{2}x
\tilde{S}^{ab}({\bf x}) K_{ab}.
\end{equation}
In contrast to the case of circular cross section, this coupling leads
to a qualitative change in the protein-protein interaction.
Proceeding as above, we find the leading distance dependence of 
protein inteaction is $1/R^{2}$:
\begin{eqnarray}
-\beta{\cal F} & = & \frac{\beta^{2}}{4} \int d^{2}x
          \int d^{2}y \tilde{S}^{ab}({\bf x}) \tilde{S}^{ij}({\bf y})  
          \langle \partial_{a}\partial_{b}h({\bf x})
          \partial_{i}\partial_{j}h({\bf y}) \rangle_{0}   \nonumber  \\
   & = & \frac{\beta^{2}}{4} \int d^{2}x
          \int d^{2}y \tilde{S}^{ab}({\bf x}) \tilde{S}^{ij}({\bf y})  
   \partial_{a}\partial_{b}\partial_{i}\partial_{j}G_{hh}({\bf x}-{\bf y})
   \nonumber  \\
   & = & \frac{1}{4} \int d^{2}x \int d^{2}y \tilde{S}^{ab}({\bf x})
         U_{ab,ij}(|{\bf x}-{\bf y}|) \tilde{S}^{ij}({\bf y}),
\end{eqnarray}
where
\begin{equation}
U_{ab,ij}(|{\bf x}-{\bf y}|) = \frac{\beta T_{abij}(\hat{\bf R})}
{2\pi \kappa |{\bf x}-{\bf y}|^{2}}.
\end{equation}
This interaction is also anisotropic, depending on ${\bf R}$ and 
$S^{ab}({\bf x}_{i})$.
For spherical cross section, since $S^{ab} \sim \delta_{ab}$ and $\delta^{ab}
\delta^{ij} T_{abij} = 0$,
the contribution to the interaction vanishes as before.
For ellipsoidal cross section, $S^{ab}=d_{1}Q^{ab}$ where $d_{1}$ is 
the coupling constant and the free energy becomes
\begin{equation}
{\cal F} = -\frac{{\cal A}^{2}d_{1}^{2}Q_{2}^{2}}{16\pi \kappa R^{2}}
\cos 2(\theta_{1}+\theta_{2}).
\end{equation}
The minimum energy configurations are at $\theta_{1}+\theta_{2} = 0, \pi$.
\par
Consequently, by introducing  symmetric-traceless tensors as the 
order-parameters for anisotropic proteins and 
by determining the relevant couplings
by symmetry, we were able to rederive the results for the circular cross
section by Goulian {\it et al.} \cite{Goul}.
Furthermore, we obtained anisotropic interactions between proteins which 
have the non-circular cross section.
This anisotropic interaction has the leading distance dependence $1/R^{4}$ and
$1/R^{2}$ depending on up-down symmetry breaking.
\setcounter{equation}{0}
\section{Model II}
In the previous section, we introduced a coupling between membrane
proteins and the height fluctuation field of the membrane by considering
symmetry and power counting. 
Since the order parameter for proteins in the coupling Eq. (\ref{pm-coup})
can be interpreted as the distribution function  of proteins,
the physical implication of this coupling can be that the bending
and Gaussian rigidities inside the protein cross section differ slightly 
from those of the surrounding membrane.
Thus this coupling can be thought of as perturbative.
However, if proteins are infinitely rigid with $\kappa = -\bar{\kappa}
= \infty$ inside the protein cross section, perturbation theory fails.
In this case, we can derive the protein-protein interaction
by considering the phenomenological interaction between membrane proteins and 
membrane lipids at the perimeter of the proteins.
\par
First, let us consider proteins which have circular cross sections
and do not break up-down symmetry.
These proteins can be modelled as inversion-symmetric three-dimensional
ellipsoids of revolution (or cylinder) with a major axis pointing along
a unit vector ${\bf m}$ in three-dimensions.
Their orientational order can be characterized by the symmetric-traceless
tensor $Q_{ij}=(m_{i}m_{j}-\delta_{ij}/3)$.
We assume the axis ${\bf m}$ prefers to align along the membrane normal
${\bf N}$.
A simple interaction favoring this alignment is
\begin{equation}
{\cal H}_{\rm int} = -\frac{1}{2}\alpha\sum_{\mu}\int_{c_{\mu}}
\frac{dl}{2\pi a_{p}}Q^{ij}N_{i}N_{j}.
\end{equation}
In the Monge gauge,
\begin{equation}
Q^{ij}N_{i}N_{j} = ({\bf m}\cdot{\bf N})^{2} - \frac{1}{3}
= - \delta^{ab}(m_{a}-N_{a})(m_{b}-N_{b}) + \mbox{constant},
\end{equation}
where $a,b$  run over 1,2  only and  $N_{a}=-\partial_{a}h$ to lowest order
in $h$.
Now we can Taylor-expand the unit normal of the membrane at the perimeter from
the center of the protein to lowest non-trivial order in $a_{p}$
\begin{equation}
N_{a}({\bf r})|_{c_{\mu}} = N_{a}({\bf R}_{\mu} + 
a_{p}{\bf b} (1-({\bf b}\cdot{\bf m}_{\perp})^{2})^{1/2})
=N_{a}({\bf R}_{\mu}) + a_{p}{\bf b}\cdot\nabla N_{a} + \cdots,
\end{equation}
where ${\bf b}$ is the unit vector from the center of the 
protein to its perimeter and $N_{a}({\bf R}_{\mu})$ is 
the average of $N_{a}({\bf r})$ along the perimeter $c_{\mu}$.
Dropping the constant term, the coupling becomes
\begin{eqnarray}
{\cal H}_{\rm int} & = & \alpha \sum_{\mu}\int_{c_{\mu}}\frac{dl}{2\pi a_{p}}
\left[ m_{a}({\bf R}_{\mu}) - N_{a}({\bf R}_{\mu}) - 
a_{p} b_{b}\partial_{b}N_{a} \right]^{2}   \nonumber  \\
   & = & \alpha \sum_{\mu} \left[ \tilde{m}_{a}({\bf R}_{\mu})
\tilde{m}_{a}({\bf R}_{\mu}) + \frac{a_{p}^{2}}{2}\partial_{b}N_{a}
\partial_{b}N_{a} \right],   
\end{eqnarray}
where $\tilde{m}_{a}({\bf R}_{\mu}) = 
m_{a}({\bf R}_{\mu}) - N_{a}({\bf R}_{\mu})$.
The free energy is
\begin{eqnarray}
e^{-\beta{\cal F}} & = & \int[{\cal D}h][{\cal D}\tilde{m}_{a}({\bf R}_{\mu})]
e^{-\beta{\cal H}_{0}-\beta{\cal H}_{\rm int}}    \\
   & = & ({\rm constant})\int[{\cal D}h] 
\exp[-\frac{1}{2}\beta\kappa\int d^{2}x
(\nabla^{2}h)^{2} - \frac{1}{2\pi}\beta\alpha {\cal A}
\sum_{\mu}\partial_{b}N_{a}({\bf R}_{\mu})
\partial_{b}N_{a}({\bf R}_{\mu})], \nonumber  
\end{eqnarray}
where ${\cal H}_{0} = \frac{1}{2}\kappa\int d^{2}x(\nabla^{2}h)^{2}$
and the integration over $\tilde{m}_{a}$ is trivial and gives
constant contribution.
Since the coupling has a quadratic form, we can evaluate this using 
the Hubbard-Stratonovich transformation. Although it is nothing
more than completing the square, we will find this technique to be
very useful.
By introducing the auxiliary fields ${\bf W}(\mu)$
and defining ${\bf V}(\mu)$  as
\begin{equation}
V_{1}(\mu) = \partial_{1}N_{1}(\mu) \;\; , \;\; V_{2}(\mu) = 
\sqrt{2}\partial_{1}N_{2}(\mu) = \sqrt{2}\partial_{2}N_{1}(\mu) \;\; , \;\; 
V_{3}(\mu) = \partial_{2}N_{2}(\mu) \;\; ,
\end{equation}
we have
\begin{eqnarray}
e^{-\beta{\cal F}} & = & \int[{\cal D}h][{\cal D}{\bf W}(\mu)] 
         \exp[ -\beta{\cal H}_{0} - \frac{1}{2\Gamma}\sum_{\mu}
         {\bf W}(\mu)\cdot{\bf W}(\mu) 
         + i\sum_{\mu}{\bf W}(\mu)\cdot{\bf V}(\mu)] \nonumber  \\
   & = & e^{-\beta{\cal F}_{0}}\int[{\cal D}{\bf W}(\mu)] 
         \exp[- \frac{1}{2\Gamma}\sum_{\mu}{\bf W}(\mu)\cdot{\bf W}(\mu)]
         \langle \exp[ i\sum_{\mu}{\bf W}(\mu)\cdot{\bf V}(\mu)] \rangle_{0},
\end{eqnarray}
where $\Gamma = \beta\alpha {\cal A}/\pi$.
Using the cumulant expansion again, to the lowest order we obtain
\begin{eqnarray}
e^{-\beta({\cal F}-{\cal F}_{0})} & = & \int[{\cal D}{\bf W}(\mu)] 
\exp[-\langle \frac{1}{2}\sum_{\mu,\nu}{\bf W}(\mu)\cdot{\bf V}(\mu)
{\bf W}(\nu)\cdot{\bf V}(\nu) \rangle_{0}]  \nonumber  \\
  & = & \int[{\cal D}{\bf W}(\mu)] \exp[-\frac{1}{2}\sum_{\mu,\nu}
W_{a}(\mu)\langle V_{a}(\mu)V_{b}(\nu)\rangle_{0}W_{b}(\nu)].
\end{eqnarray}
For two proteins separated by a distance $R$, we find
\begin{eqnarray}
e^{-\beta({\cal F}-{\cal F}_{0})} & = & \int[{\cal D}{\bf W}(1)]
[{\cal D}{\bf W}(2)] \exp[-\frac{1}{2}\sum_{\mu,\nu=1}^{2}
W_{a}(\mu)\langle V_{a}(\mu)V_{b}(\nu)\rangle_{0}W_{b}(\nu)]  \nonumber  \\
   & = & \left( \det \langle V_{a}(\mu)V_{b}(\nu)\rangle_{0} \right)^{-1/2}  
         \nonumber  \\
   & = & ({\rm constant})
         \left( 1-3(\frac{8}{R^{2}\Lambda^{2}})^{2}\right)^{-1/2},
\end{eqnarray}
where $\Lambda \sim 1/a_{p}$ is a cut-off for the height fluctuation,
and $a_{p}$ is the radius of the protein introduced in Sec. II.A.
Thus, the free energy has an $R$-dependence as
\begin{equation}
{\cal F} = - k_{B}T\frac{96}{R^{4}\Lambda^{4}} = 
- k_{B}T\frac{6{\cal A}^{2}}{\pi^{2}R^{4}} ,
\label{eq:alpha=0}
\end{equation}
in accord with the previous calculation by Goulian {\it et al.} \cite{Goul}.
In the above equation, we used a cut-off for the height fluctuation,
$\Lambda = 2/a_{p}$ \cite{cutoff}.
\par
For proteins that break the up-down symmetry, the unit normal of the 
membrane at the perimeter of the protein is not forced to be parallel
to the direction of the protein. Instead, the unit normal is forced to have
a fixed angle $\alpha_{\mu}$ with the  direction of the $\mu$-th protein. Thus
the preferred unit normal at the perimeter is 
\begin{equation}
{\bf N}_{0}({\bf r})|_{c_{i}} = \frac{{\bf m}+\alpha_{\mu}{\bf b}}
{\sqrt{1+\alpha_{\mu}^{2}}}.
\end{equation}
The coupling between the protein and the membrane at the perimeter of the
protein is
\begin{equation}
{\cal H}_{\rm int}= \alpha\sum_{\mu}\int_{c_{\mu}}\frac{dl}{2\pi a_{p}}
({\bf N}-{\bf N}_{0})^{2}.
\end{equation}
We Taylor expand to find to lowest order 
\begin{eqnarray}
{\cal H}_{\rm int} & = & \alpha\sum_{\mu}\int_{c_{\mu}}\frac{dl}{2\pi a_{p}}
[ (m_{a}-N_{a})(m_{a}-N_{a}) - 2\alpha_{\mu}N_{a}b_{a} ]   \nonumber  \\
   & = & \alpha\sum_{\mu} [ \tilde{m}_{a}(\mu)\tilde{m}_{a}(\mu) +
\frac{1}{2}a_{p}^{2}\partial_{b}N_{a}\partial_{b}N_{a} -
a_{p}\alpha_{\mu}\partial_{a}N_{a} ] .
\end{eqnarray}
The free energy is now
\begin{eqnarray}
e^{-\beta{\cal F}} & = & \int[{\cal D}h][{\cal D}\tilde{m}_{a}(\mu)]
\exp [ -\frac{1}{2}\beta\kappa\int d^{2}x(\nabla^{2}h)^{2}   \nonumber  \\
     &   &  - \beta\alpha\sum_{\mu} [ \tilde{m}_{a}(\mu)\tilde{m}_{a}(\mu) +
\frac{1}{2}a_{p}^{2}\partial_{b}N_{a}\partial_{b}N_{a} -
a_{p}\alpha_{\mu}\partial_{a}N_{a} ]  ]                             \\
   & = & \int[{\cal D}h] \exp\left[ -\frac{1}{2}\beta\kappa
\int d^{2}x(\nabla^{2}h)^{2} -\frac{\beta\alpha {\cal A}}{2\pi}
\sum_{\mu}[\partial_{b}N_{a}\partial_{b}N_{a} -
\frac{2\alpha_{\mu}}{a_{p}} \partial_{a}N_{a} ]  \right]   . \nonumber
\end{eqnarray}
Re-defining ${\bf V}(\mu)$ as
\begin{equation}
V_{1}(\mu) = \partial_{1}N_{1}(\mu) - \frac{\alpha_{\mu}}{a_{p}} \;\; , \;\;
V_{2}(\mu) = \partial_{1}N_{2}(\mu) = \partial_{2}N_{1}(\mu) \;\; , \;\;
V_{3}(\mu) = \partial_{2}N_{2}(\mu) - \frac{\alpha_{\mu}}{a_{p}} \;\; ,
\end{equation}
the free energy becomes
\begin{eqnarray}
e^{-\beta{\cal F}} & = & \int[{\cal D}h]
       \exp\left[ -\frac{1}{2}\beta\kappa
       \int d^{2}x(\nabla^{2}h)^{2} - \frac{\beta\alpha {\cal A}}  
       {2\pi}\sum_{\mu} {\bf V}(\mu)\cdot{\bf V}(\mu) \right]  \nonumber  \\
   & = & \int[{\cal D}h][{\cal D}{\bf W}(\mu)]
       \exp\left[ -\frac{1}{2}\beta\kappa
       \int d^{2}x(\nabla^{2}h)^{2} - \frac{1}{2\Gamma}
       \sum_{\mu} {\bf W}(\mu)\cdot{\bf W}(\mu) + 
       i\sum_{\mu}{\bf W}(\mu)\cdot{\bf V}(\mu) \right]  \nonumber  \\
   & = & \int[{\cal D}{\bf W}(\mu)]\exp[- \frac{1}{2\Gamma}
       \sum_{\mu} {\bf W}(\mu)\cdot{\bf W}(\mu) ] 
       \langle \exp[i\sum_{\mu}{\bf W}(\mu)\cdot{\bf V}(\mu) ] \rangle_{0},
\end{eqnarray}
where $\Gamma=\beta\alpha{\cal A}/\pi$.
For two proteins separated by a distance $R$, we find
\begin{eqnarray}
e^{-\beta{\cal F}} & = & \int[{\cal D}{\bf W}(1)][{\cal D}{\bf W}(2)]
   \exp[ i\sum_{\mu=1}^{2}{\bf W}(\mu)\cdot\langle {\bf V}(\mu)\rangle_{0}  
         \nonumber  \\
   &   & -\frac{1}{2}\sum_{\mu,\nu=1}^{2}
          \left( \langle {\bf W}(\mu)\cdot{\bf V}(\mu)
         {\bf W}(\nu)\cdot{\bf V}(\nu) \rangle_{0} - 
         \langle {\bf W}(\mu)\cdot{\bf V}(\mu) \rangle_{0} 
         \langle {\bf W}(\nu)\cdot{\bf V}(\nu) \rangle_{0} \right) ] 
         \nonumber  \\
   & = & \int[{\cal D}\tilde{W}] e^{-\frac{1}{2}\tilde{W}\tilde{M}\tilde{W}
         -i\tilde{A}\tilde{W}}   \nonumber  \\
   & = & (\det\tilde{M})^{-1/2} 
         e^{-\frac{1}{2}\tilde{A}\tilde{M}^{-1}\tilde{A}},
\end{eqnarray}
where
\begin{eqnarray}
\tilde{W} &=& (W_{1}(1), W_{2}(1), W_{3}(1), W_{1}(2), W_{2}(2), W_{3}(2)), \\
\tilde{A} &=& (\frac{\alpha_{1}}{a_{p}}, 0, \frac{\alpha_{1}}{a_{p}},
               \frac{\alpha_{2}}{a_{p}}, 0, \frac{\alpha_{2}}{a_{p}}),   \\
\tilde{M} & = & \frac{\Lambda^{2}}{32\pi\beta\kappa}
                \left( \begin{array}{cccccc}
                3 & 0 & 1 & -\lambda & 0 & -\lambda  \\
                0 & 2 & 0 & 0 & -2\lambda & 0        \\
                1 & 0 & 3 & -\lambda & 0 & 3\lambda  \\
                -\lambda & 0 & -\lambda & 3 & 0 & 1  \\
                0 & -2\lambda & 0 & 0 & 2 & 0        \\
                -\lambda & 0 & 3\lambda & 1 & 0 & 3  
                \end{array}
                \right)
\end{eqnarray}
with $\lambda = 8/R^{2}\Lambda^{2}$ and $\Lambda = 2/a_{p}$ is the 
membrane cutoff.
Thus the $R$-dependence of the free energy becomes
\begin{eqnarray}
-\beta{\cal F} & = & -\frac{1}{2}\ln(1-3(\frac{8}{R^{2}\Lambda^{2}})^{2})
-\frac{1}{2}\frac{16\pi\beta\kappa}{\Lambda^{2}}
(1+\frac{1}{2}(\frac{8}{R^{2}\Lambda^{2}})^{2})
(\alpha_{1}^{2}+\alpha_{2}^{2})/a_{p}^{2}  \nonumber  \\
   & = & \frac{3}{2}(\frac{8}{R^{2}\Lambda^{2}})^{2} -
         \frac{4\cdot 64\pi\beta\kappa}{a_{p}^{2}\Lambda^{6}}
         \frac{\alpha_{1}^{2}+\alpha_{2}^{2}}{R^{4}}.
\end{eqnarray}
Our final form for the free energy is
\begin{equation}
{\cal F} = -k_{B}T\frac{6{\cal A}^{2}}{\pi^{2}R^{4}} +
\frac{4\kappa {\cal A}^{2}}{\pi}
\frac{\alpha_{1}^{2}+\alpha_{2}^{2}}{R^{4}}.
\label{al-not-0}
\end{equation}
This gives the previous result Eq.~(\ref{eq:alpha=0}) for $\alpha_{\mu}=0$
which corresponds to the strong-coupling regime in Ref. \cite{Goul}. 
In the limit $T \rightarrow 0$, this gives the result for the low temperature
regime in Ref. \cite{Goul}.
Thus, in this phenomenological model, we obtain the general interaction between
the up-down symmetry breaking proteins at finite temperature $T$.
\par
This calculation, which focuses on the change in free energy brought about
by the addition of inclusions, does not show explicitly how these inclusions
modify the shape of the membrane at large distances from the inclusions.
Careful treatment of the minimum energy configuration of $h$, about which
we calculated Gaussian fluctuations, yields the same large distance distortion
$(h \sim \cos n\theta)$ as calculated by Goulian {\it et. al.}
We believe this result to be true for a free membrane with no imposed 
boundary conditions.
If the membrane is forced to be flat by an aligning field, these long-range 
forces will become short-range and surface tension will have a similar effect.
It is not so clear what will happen on a vesicle of spherical topology
with no Laplace pressure.
This question is currently under investigation.
\setcounter{equation}{0}
\section{Height-Displacement Model}
Non-transmembrane proteins are exposed to a specific surface of a membrane. 
Thus, they have preferred 
center-of-mass positions not at the center of the bilayer 
(See fig. \ref{fig4}). 
We consider the interaction between these proteins 
by introducing the potential energy $\Psi(\zeta-h)$ where 
$\zeta$ is the position
of the protein and $h$ is the membrane height fluctuation field.
For integral proteins, $\Psi(\zeta-h)$ has a minimum at the non-vanishing
value of $\zeta-h = r_{0}$. We can expand $\Psi(\zeta-h)$ in terms
of the deviation from this preferred value $r_{0}$
\begin{equation}
\Psi(\zeta-h)=\Psi(r_{0}) +\frac{1}{2}((\zeta-h)-r_{0})^{2}
\frac{\partial^{2}\Psi}{\partial\zeta^{2}}|_{(\zeta-h)=r_{0}} + \cdots,
\end{equation}
and if proteins are tightly bound, $\frac{\partial^{2}\Psi}
{\partial\zeta^{2}}|_{(\zeta-h)=r_{0}} \gg 1$.
Considering the symmetry, we introduce the couplings
\begin{eqnarray}
{\cal H}_{\rm int} & = & \frac{1}{2}\sum_{\mu}\int_{D_{\mu}}d^{2}x [
         k_{1}({\bf m}+\nabla h)^{2} + k_{2}(\zeta-h)\nabla^{2}h  \nonumber  \\
   &   & + k_{3}(\zeta-h)m_{a}m_{b}\partial_{a}\partial_{b}h
         + \Psi(\zeta-h) ],
\end{eqnarray}
where ${\bf m}$ is the preferred direction of the protein.
By minimizing over $m_{a}$, we find
\begin{equation}
m_{a} = -\partial_{a}h + \frac{k_{3}}{k_{1}}(\zeta-h)\partial_{b}h
\partial_{a}\partial_{b}h + {\cal O}(h^{3}).
\end{equation}
Substituting this result into the coupling,
we obtain in lowest order
\begin{eqnarray}
{\cal H}_{\rm int} & = & \frac{1}{2}\sum_{\mu}\int_{D_{\mu}}d^{2}x [
k_{2}(\zeta-h)\nabla^{2}h  + 
k_{3}(\zeta-h)\partial_{a}h\partial_{b}h\partial_{a}\partial_{b}h  
\nonumber  \\
   &   & + \Psi(r_{0}) +\frac{1}{2}((\zeta-h)-r_{0})^{2}
\frac{\partial^{2}\Psi}{\partial\zeta^{2}}|_{(\zeta-h)=r_{0}} ].
\end{eqnarray}
Minimizing over $(\zeta-h)$ gives the preferred position of the protein as
\begin{equation}
\zeta-h = r_{0} - k_{2}
\left(\frac{\partial^{2}\Psi}{\partial\zeta^{2}}|_{(\zeta-h)=r_{0}} 
\right)^{-1} \nabla^{2}h + {\cal O}(h^{3}).
\end{equation}
Thus we obtain the coupling
\begin{equation}
{\cal H}_{\rm int} =  \frac{1}{2}\sum_{\mu}\int_{D_{\mu}}d^{2}x [
k_{2}r_{0}\nabla^{2}h + k_{3}r_{0}\partial_{a}h\partial_{b}h
\partial_{a}\partial_{b}h -\frac{1}{2}k_{2}^{2}
\left(\frac{\partial^{2}\Psi}{\partial\zeta^{2}}|_{(\zeta-h)=r_{0}} 
\right)^{-1} (\nabla^{2}h)^{2} ].
\label{eq:k2=0}
\end{equation}
The first and the last terms look similar to the ones in the 
phenomenological model.
However, the non-linear second term is allowed because the up-down
symmetry is broken by the preferred position of the protein.
In the low temperature limit, we assume lipids are so tightly bound to
the proteins that $\left(\frac{\partial^{2}\Psi}
{\partial\zeta^{2}}|_{(\zeta-h)=r_{0}} \right)$ is
much bigger than $k_{2}$ and we drop the last term in Eq. (\ref{eq:k2=0}).
In this limit, the Hamiltonian becomes
\begin{equation}
{\cal H} = \frac{1}{2}\kappa\int d^{2}x (\nabla^{2}h)^{2}
+ \frac{1}{2}\sum_{\mu}\int_{D_{\mu}}d^{2}x [
k_{2}r_{0}\nabla^{2}h + k_{3}r_{0}\partial_{a}h\partial_{b}h
\partial_{a}\partial_{b}h ].
\end{equation}
Now by minimizing this Hamiltonian over $h$,
we obtain the low temperature limit for the protein interactions.
From the minimum condition,
\begin{eqnarray}
\frac{\delta{\cal H}}{\delta h} = 0 & = & \kappa \nabla^{4}h +
\sum_{\mu}k_{2}r_{0}\nabla^{2}\delta({\bf r}-{\bf r}_{\mu})   \nonumber  \\
   &   & + \sum_{\mu}k_{3}r_{0} (2\partial_{a}\delta({\bf r}-{\bf r}_{\mu})
\partial_{b}h\partial_{a}\partial_{b}h + \partial_{a}h\partial_{b}h
\partial_{a}\partial_{b}\delta({\bf r}-{\bf r}_{\mu})),
\end{eqnarray}
we obtain the equilibrium height for the membrane
\begin{equation}
h(r) = \sum_{\mu}G({\bf r}-{\bf r}_{\mu}) -
\frac{9k_{3}}{k_{2}^{2}r_{0}}\int d^{2}r' G({\bf r}-{\bf r}')
\sum_{\mu,\nu,\lambda}\partial_{a}G({\bf r}'-{\bf r}_{\mu})
\partial_{b}G({\bf r}'-{\bf r}_{\nu})
\partial_{a}\partial_{b}G({\bf r}'-{\bf r}_{\lambda}).
\end{equation}
Thus the interaction becomes 
\begin{eqnarray}
{\cal H} & = & - \frac{1}{2}\kappa\int d^{2}x 
   \left( \sum_{\mu}\nabla^{2}G({\bf r}-{\bf r}_{\mu}) + \frac{9k_{3}}{k_{2}}
   \sum_{\mu,\nu,\lambda}\partial_{a}G({\bf r}-{\bf r}_{\mu})
    \partial_{b}G({\bf r}-{\bf r}_{\nu})
 \partial_{a}\partial_{b}G({\bf r}-{\bf r}_{\lambda})\right)^{2} \nonumber  \\
   & \simeq & -\frac{1}{2}\kappa \sum_{\mu,\nu}
   \left(\frac{k_{2}r_{0}}{\kappa}\right)^{2}
   \delta({\bf r}_{\mu}-{\bf r}_{\nu})  \nonumber  \\
   &   & - \kappa \sum_{\mu,\nu,\lambda,\sigma}
   \left( -\frac{k_{2}r_{0}}{\kappa}\right)
   \frac{9k_{3}}{k_{2}}\partial_{a}G({\bf r}_{\mu}-{\bf r}_{\nu})
   \partial_{b}G({\bf r}_{\mu}-{\bf r}_{\lambda})
   \partial_{a}\partial_{b}G({\bf r}_{\mu}-{\bf r}_{\sigma})   .
\label{eq:gen-low}             
\end{eqnarray}
For two proteins separated by a distance $R$, we obtain the leading 
distance dependence of the interaction 
\begin{eqnarray}
{\cal H} & = & 18k_{3}r_{0}\partial_{a}G(R)\partial_{b}G(R)
   \partial_{a}\partial_{b}G(R)      \nonumber  \\           
   & = & -18k_{3}r_{0} \left( \frac{k_{2}r_{0}}{2\pi\kappa} \right)^{3}
   \frac{R_{a}R_{b} (\delta_{ab}R^{2} - 2R_{a}R_{b})}{R^{8}}  \nonumber  \\
   & = & \left( \frac{k_{2}r_{0}}{2\pi\kappa} \right)^{3}
   \frac{18k_{3}r_{0}}{R^{4}}.
\label{eq:low-lim}
\end{eqnarray}
Above we used $G(R) = -(k_{2}r_{0}/4\pi\kappa)\ln R^{2}$.
We can interprete the parameters in Eq. (\ref{eq:low-lim}) as the
area of proteins and the contact angle between proteins and lipids.
 
When there are several proteins, from Eq. (\ref{eq:gen-low}), 
we find that three- and four-body interactions exist in addition to 
two-body interaction.
For three proteins separated by ${\bf R}_{\mu\nu}$ which is the vector from
the $\nu$-th protein to the $\mu$-th protein, 
we find three-body interaction to be 
\begin{equation}
{\cal H}_{\rm 3-body} = - 18k_{3}r_{0}
    \left( \frac{k_{2}r_{0}}{4\pi\kappa} \right)^{3}
    \sum'_{\mu,\nu,\sigma}\frac{1}{R_{\mu\nu}^{2}R_{\mu\sigma}^{2}} 
      \left( 1-\frac{2({\bf R}_{\mu\nu}\cdot {\bf R}_{\mu\sigma})^{2}}
      {R_{\mu\nu}^{2}R_{\mu\sigma}^{2}}
      -\frac{2{\bf R}_{\mu\nu}\cdot {\bf R}_{\mu\sigma}}{R_{\mu\nu}^{2}}
      \right),
\end{equation}
where $\sum'$ means all $\mu,\nu,\sigma$ are different,
in addition to two-body interaction between each pair of proteins given by
Eq. (\ref{eq:low-lim}).
Similarly, we find four-body interaction to be
\begin{equation}
{\cal H}_{\rm 4-body} = - 9k_{3}r_{0}
        \left( \frac{k_{2}r_{0}}{4\pi\kappa} \right)^{3}
        \sum'_{\mu,\nu,\lambda,\sigma}
        \frac{1}{R_{\mu\nu}^{2}R_{\mu\lambda}^{2}R_{\mu\sigma}^{2}} 
   \left( {\bf R}_{\mu\nu}\cdot {\bf R}_{\mu\lambda} 
   - \frac{2{\bf R}_{\mu\nu}\cdot {\bf R}_{\mu\lambda} 
             {\bf R}_{\mu\lambda}\cdot {\bf R}_{\mu\sigma}}
       {R_{\mu\lambda}^{2}}
           \right),  
\end{equation}
where $\sum'$ means all $\mu,\nu,\lambda,\sigma$ are different.
Note that these three- and four-body interactions are also $1/R^{4}$
interaction which is the same order as two-body interaction.
\section{Discussion}
We model biological membrane as a continuous bilayer of
lipid molecules in which various membrane inclusions such as proteins
are embedded. 
Such model membranes with inclusions also have potetial applications for
target drug delivery, nano-scale pumps, functionized interfaces, and
chemical reactors.
In this paper, we study how the membrane contributes to the interactions
between inclusions.
Also, it is interesting to understand how inclusions affect the properties 
such as rigidity or shape of
model membranes.
\par
The interaction between membrane inclusions such as proteins with
circular cross-sectional area was first calculated by Goulian {\it et al.}
Using three models, which we refer to as Model I, Model II,
and a height-displacement models, 
we recover all the results by Goulian {\it et al.} 
The interaction in Eq. (2.2) is a temperature-dependent interaction between
two circular inclusions that falls off with distance as $1/R^{4}$.
Assuming two inclusions are identical, the force will be attractive if
$\delta\kappa$ and $\delta\bar{\kappa}$ have the opposite sign and repulsive 
otherwise.
%In particular, for hard inclusions, $\delta\kappa > 0$ and $\delta\bar{\kappa}
%< 0$. 
%Consequently, the force is attractive between two hard inclusions.
%In Eq. (2.5), we find the consistent result with the one described above
%between two hard inclusions using different model.
For inclusions with up-down symmetry, the interaction is attractive and
falls off as $1/R^{4}$ again.
The magnitude is set by $k_{B}T$ and is independent of the rigidity $\kappa$.
When inclusions break up-down symmetry, 
in addition to the attractive interaction set by $k_{B}T$,
we find a repulsive interaction proportional to the square of the contact
angle in Eq. (2.6).
Thus, for up-down asymmetric inclusions, there are competing attractive and 
repulsive interactions and we might have an interesting transition between
aggregation and mixing of inclusions when $k_{B}T \sim \kappa\alpha^{2}$.
Both Model I (for soft inclusions) and Model II (for hard inclusions)
predict potentials that fall off with distance as $R^{-4}$.
This interaction is attractive for hard inclusions.
For soft inclusions the sign of the interaction depends on the relative
sign of $\delta \kappa$ and $\delta\bar{\kappa}$ and is attractive if
they have opposite signs.
Reasonable models predict $\delta\kappa\delta\bar{\kappa} < 0$
so that the prediction of models I and II can be viewed as being consistent.
\par
Furthermore, we calculate the interaction between proteins with 
the non-circular cross-sectional area and find anisotropic $1/R^{2}$ and 
$1/R^{4}$ interactions depending on whether up-down symmetry is broken
or not.
In Eq. (2.3), we find the interaction between proteins with non-circular
cross-sectional area when up-down symmetry is conserved, and the free
energy again falls off as $1/R^{4}$ but the magnitude depends on the
orientations of inclusions.
When only 4-fold anisotropy is nonvanishing, the angular dependence of
the interaction is of the form $\cos 4(\theta_{1}+\theta_{2})$.
The interaction depends on the relative orientations of inclusions
to the separation vector from one inclusion and the other.
The interaction is attractive if $|\theta_{1}+\theta_{2}| < \pi/8$
and repulsive if $\pi/8 < |\theta_{1}+\theta_{2}| < \pi/4$.
Thus, depending on the orientations of the inclusions with respect to the
separation vector, the interaction between two non-circular inclusions can
be attractive or repulsive.
In general, 2-fold and 4-fold anisotropies are nonvanishing and the 
resulting interaction has more complicated orientational dependence as shown
in Eq. (2.3).
In the case of broken up-down symmetry, 
the interesting aspect of the interaction in Eq. (2.4), in addition to
the orientational dependence of the force, is the leading distance-dependence
$1/R^{2}$ rather than $1/R^{4}$.
The angular dependence in this case is of the form 
$\cos 2(\theta_{1}+\theta_{2})$ and the interaction is attractive if
$|\theta_{1}+\theta_{2}| < \pi/4$
and repulsive if $\pi/4 < |\theta_{1}+\theta_{2}| < \pi/2$.
In general, transmembrane proteins are asymmetrically embedded in the
membrane, and they break up-down symmetry of bilayer membranes.
Thus, these proteins interact with anisotropic $1/R^{2}$ interaction,
which is much stronger at large length-scale than the screened electrostatic
interaction or the Van der Waals interaction under physiological conditions.
Consequently, for a distance large compared  with a typical protein size,
the interaction described in this paper will dominate over the electrostatic
and the Van der Waals interactions.
\par
Recently, we received the preprint by Golestanian {\it et al.} \cite{Goul2}
in which 
they extend the calculation of Ref. \cite{Goul} to the interaction between
two rods on membranes. In their work, they retain the up-down symmetry for
the rods and obtain anisotropic $1/R^{4}$ interaction 
similar to Eq~(\ref{quad-int}).
Also, we'd like to mention the calculation of the short-ranged 
induced interactions between inclusions embedded in fluid membrane 
by Dan {\it et al.} \cite{Dan,Dan2}. 
%They find that taking into account the bilayer
%bending rigidity leads to an oscillatory decay in the membrane thickness
%with distance from the inclusion boundary at short length-scale.
%As a result, the short-range, membrane-induced interactions between inclusions
%are non-monotonic.
They find a short-range repulsive interaction with decaying oscillation
with period of order the membrane thickness when the two halves of
a bilayer membrane are allowed to respond separately to the
membrane inclusions.
Consequently, they suggest that in systems where the inclusions impose specific
contact angles, a meta-stable state with a well-defined separation between 
neighboring inclusions is possible and the minimal energy state of 
the membrane is obtained at a finite inclusion spacing.
%With this short-range interaction and the long-range interaction described 
%in this paper, membrane proteins aggregate at large distance compared with 
%their size and at the distance scale comparable to their size, 
%they keep a finite inclusion spacing.
The competition between this short-range repulsive force and the long-range
attractive forces discussed by Goulian {\it et. al.} and in this paper
could lead to a preferred separation between membrane proteins.
Thus, it may be suggested that a hydrophilic channel through the bilayer 
can be formed by a ring of three or more transmembrane proteins, 
which may give some idea about how protein molecules can facilitate 
the passage of ions or molecules into and out of cells.
\par
We are grateful to Mark Goulian for helpful discussions and for pointing 
out an error in our treatment of the cutoff in an early version of
this manuscript. 
This work was supported in part by the National Science Foundation under
grant No. DMR94-23114 and by the Penn Laboratory for Research in the
Structure of Matter under NSF grant No. DMR91-20668.

\clearpage
\input{psfig}
\input{psfig}
\input{psfig}
\begin{figure}
\centerline{\psfig{figure=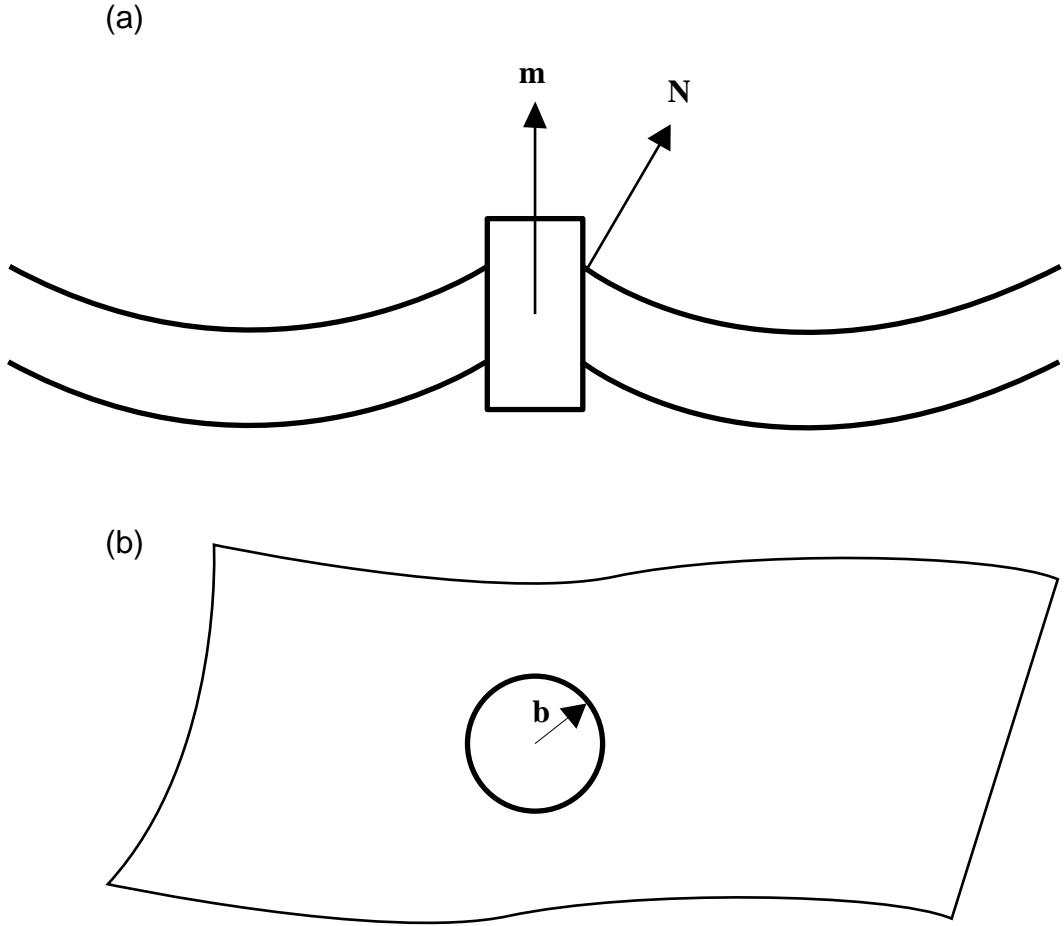}}
\caption{(a) The unit vector ${\bf m}$ denotes the direction of 
the protein and ${\bf N}$ denotes the unit normal vector of the membrane 
at the perimeter of protein (Side view). (b) ${\bf b}$ is the unit vector
from the center of the protein to its perimeter (Top view).
}
\label{fig3}
\end{figure}
\input{psfig}
\begin{figure}
\centerline{\psfig{figure=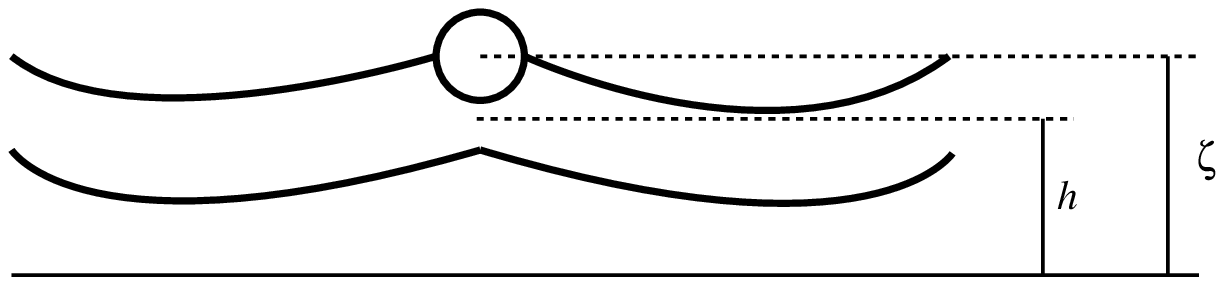}}
\caption{Non-transmembrane proteins have preferred center-of-mass positions
nat at the center of the bilayer. $\zeta$ denotes the position of the protein
and $h$ is the membrane height fluctuation field.
The potential energy has a minimum at the non-vanishing value of
$\zeta-h=r_{0}$.
}
\label{fig4}
\end{figure}


\begin{thebibliography}{99}
\bibitem{cutoff} To calculate a cut-off $\Lambda$ for the height
fluctuation approximately, we use the fact that the number of points in the
first Brillouin zone is equal to the number of sites in the real space.
Since the height fluctuation is forbidden on scales less than the radius of
the proteins, we find
$$
\frac{\pi\Lambda^{2}}{(2\pi/L)^{2}} = \frac{L^{2}}{\pi a_{p}^{2}}, \nonumber
$$
where $L$ is the linear size of the membrane.
Thus, we obtain $\Lambda=2/a_{p}$ and this approximation is consistent
with the ultra-violet regularization adapted in Ref. \cite{Goul}.
\end{thebibliography}
\end{document}